\documentclass[a4paper,12pt]{article}
\usepackage{amsthm,amsmath,amssymb}

\usepackage{setspace}

\usepackage[english]{babel}
\usepackage[margin=1in]{geometry}

\usepackage{xcolor}
\usepackage{natbib}
\theoremstyle{plain}

\theoremstyle{definition}
\usepackage[]{graphicx}
\chardef\bslash=`\\ 

\usepackage{authblk}

\title{Variable selection in linear regression models: choosing the best subset is not always the best choice}

\author[1]{Moritz Hanke\footnote{Shared first authorship}\footnote{Corresponding author. E-mail: \texttt{hanke@leibniz-bips.de}}}
\author[1]{Louis Dijkstra$^\ast$}
\author[1]{Ronja Foraita}
\author[1]{Vanessa Didelez}

\affil{Leibniz Institute for Prevention Research \& Epidemiology -- BIPS\\  Achterstr. 30, 28359 Bremen (Germany)}

\date{December 2022}                     

\begin{document}

\maketitle 

\begin{abstract}
\noindent Variable selection in linear regression settings is a much discussed problem. Best subset selection (BSS) is often considered the intuitive 'gold standard’, with its use being restricted only by its NP-hard nature. Alternatives such as the least absolute shrinkage and selection operator (Lasso) or the elastic net (Enet) have become methods of choice in high-dimensional settings. A recent proposal represents BSS as a mixed integer optimization problem so that much larger problems have become feasible in reasonable computation time. We present an extensive neutral comparison assessing the {\em variable selection} performance, in linear regressions, of BSS compared to forward stepwise selection (FSS), Lasso and Enet. The simulation study considers a wide range of settings that are challenging with regard to dimensionality (with respect to the number of observations and variables), signal-to-noise ratios and correlations between predictors. As main measure of performance, we used the best possible F1-score for each method to ensure a fair comparison irrespective of any criterion for choosing the tuning parameters, and results were confirmed by alternative performance measures. 
Somewhat surprisingly, it was {\em only} in settings where the signal-to-noise ratio was high and the variables were (nearly) uncorrelated that BSS reliably outperformed the other methods, even in  low-dimensional settings. Further, the FSS’s performance was nearly identical to BSS. Our results shed new light on the usual presumption of BSS being, in principle, the best choice for variable selection. Especially for correlated variables, alternatives like Enet are faster and appear to perform better in practical settings. \\[.2cm]
\end{abstract}






\section{Introduction}

Selecting a subset of variables as direct predictors for an outcome is a much studied problem in (linear) regression modelling and has received renewed attention in the context of high-dimensional data where some variable selection is unavoidable. It might appear self-evident that best subset selection (BSS) \citep{Beale_1967,Hocking_1967,Garside_1971} should be the gold standard for variable selection: Clearly, if we assume there are $s$ direct predictors and consider \textit{all} combinations of variables up to a subset size $k \geq s$ the true model has to be one of the candidate models, making BSS the obvious choice for variable selection. The main reason for dismissing BSS is that a naive implementation quickly becomes computationally infeasible\footnote{For example, with $p=100$ and a subset size $k=15$ there are over $2.533 \cdot 10^{17}$ possible sets.} with larger numbers of variables \citep{Hastie_2020}. In their path-breaking work, \citet{Bertsimas_2016} have formulated BSS as a mixed integer optimization problem (MIO) pushing the boundaries for the feasible number of variables $p$ to be in the thousands while still searching over moderate subset sizes of $k$. 
This now allows a better comparison of BSS with variable selection methods such as the popular Least absolute shrinkage estimator (Lasso) \citep{Tibshirani_1996} or variants thereof, e.g. the adaptive Lasso \citep{Zou_2006} or the Elastic net (Enet) and its adaptive version \citep{Zou_2005,Zou_2009}. 
The convex optimization nature of all these methods enables quick computation even for millions of variables making them the main methods of choice in high-dimensional settings \citep{Qian_2020}. While these methods' theoretical and empirical performances have been studied in much detail \citep{Hastie_2009,Chun_2010,Buehlmann_2011,Xu_2012,Houwelingen_2013,Sanchez_2018,Yu_2019,Lima_2020,Wang2020,Lederer_2022}, the question arises how they actually compare empirically to BSS in realistic high-dimensional settings. This is of much interest especially in the fields of genetics and bioinformatics, where high-dimensional selection problems are common and $L_1$-, $L_2$- or $L_\infty$-penalisation approaches are often used as computationally feasible alternatives to BSS \citep{Li_2018,Atyeo_2021,Overmyer_2021}.

A first extensive comparison of BSS with the Lasso and a simplified version of the relaxed Lasso \citep{Meinshausen_2007} regarding their predictive performance has been carried out by \citet{Hastie_2020}. Perhaps surprisingly, the authors find that neither BSS nor the Lasso uniformly dominate each other, and, moreover, that forward step-wise selection is not worse than BSS, while the relaxed Lasso shows the best performance overall. An explanation may be found in the different bias-variance trade-offs of the different approaches which enter all the metrics for assessing predictive accuracy considered by \citet{Hastie_2020}.
Their results may, therefore, not hold up for variable selection performance, since different sets of selected variables can give very similar predictions, but only one set of variables is the true set of direct predictors. 
Another recent simulation study using the MIO formulation was carried out by \citet{Takano_2020}. The authors compared BSS with Lasso based on different optimization criteria determining the subset size $k$. However, in their study, they only considered a low dimensional setting with $n=100$ observations and $p=25$ predictors. 

In the present paper, we complement the above studies by specifically evaluating the selection performance of BSS compared to other established approaches, gaining further insights into the properties of all these methods. Selecting the true direct predictors of an outcome, relative to a large set of available variables, is a distinct problem from prediction and of much substantial interest in its own right, for instance in genetics. We take advantage of the MIO formulation, which allows BSS to be used in high-dimensional settings. 
While we choose a similar setting for our simulation study as \citet{Hastie_2020}, we also extend their approach in several important ways: 
For wider applicability, we consider more complex situations than just the Toeplitz correlation structure; 
we also choose different positions of the direct predictors within those correlation structures. For added realism, we supplement the fully-synthetic simulations with semi-synthetic simulations where we use the actually observed correlation structure from real data, namely gene expressions provided by The Cancer Genome Atlas Program \citep{TCGA2011}. 
Overall our simulation has 276 different parameter combinations and the results can be compared in an interactive web-app at {\tt{ https://bestsubset.bips.eu}}.

All methods considered in the present paper require choosing a tuning parameter or subset size potentially affecting which and how many variables are selected. To enable a neutral and fair comparison, we choose for each method its optimal tuning parameter (or subset size) in terms of its best achievable F1-score.
This allows us to assess the best \textit{possible} variable selection performance, separating this from the issue of choosing a tuning parameter. In practice, there appears to  be no gold-standard for choosing the tuning parameter, especially in high-dimensions. Instead, to ensure a fair and practically feasible comparison of the methods we use an alternative approach by choosing the tuning parameters to obtain a given number of selected variables, i.e. subset size.

The paper is organized as follows: The methods section describes the selection procedures under investigation and their theoretical properties. Subsequently, we describe the setup of the simulation study as well as our findings. Finally, we will draw conclusions and give an outlook on future work in the last section. 

\section{Methods}
Given a vector of responses $\mathbf{y} \in \mathbb{R}^n$, a design matrix $\mathbf{X} \in \mathbb{R}^{n \times p}$, a vector of coefficients $\boldsymbol{\beta} \in \mathbb{R}^p$ and a noise vector $\boldsymbol{\epsilon} \in \mathbb{R}^n$ with $\epsilon_i \sim \mathcal{N}(0, \sigma^2)$ for independent $i=1, \dots, n$, we assume the linear model 
\begin{align}\label{eq:lin_reg_model}
\mathbf{y} = \mathbf{X}\boldsymbol{\beta} + \boldsymbol{\epsilon}
\end{align}
where $\mathbf{x}_{j}$, $j=1, \dots, p$, has been standardized such that $\sum_{i=1}^{n}x_{i,j}=0$ and $n^{-1}\sum_{i=1}^{n}x_{i,j}^2=1$. We further assume $\boldsymbol{\beta}$ to be sparse in the sense that for $s = \sum_{j=1}^{p}I(\beta_j \neq 0)$ we have $s = \mathcal{O}(n^c)$ for $ 0 < c < 1$ \citep{Meinshausen_2006,Zhao_2006}. 
Let $\boldsymbol{\hat{\beta}}$ denote an estimator of $\boldsymbol{\beta}$ and $\text{supp}(\cdot)$ the support of a vector, i.e. indicating which elements are non-zero. 
	We say that an estimator or procedure is selection consistent if $\text{supp}(\hat{\boldsymbol{\beta}})$ converges in probability to $\text{supp}(\boldsymbol{\beta})$. 

While the ordinary least squares (OLS) estimator is the best linear unbiased estimator for $\boldsymbol{\beta}$ when $n>p$, it is not suited for variable selection where the aim is to discriminate zeros from non-zeros in $\boldsymbol{\beta}$. For $\beta_j =0$, it can be shown that the OLS estimate is $\hat{\beta}_j = \mathcal{O}\left(\sqrt{n^{-1} \log n}\right)$ \citep{Horowitz_2015}, i.e., it does not select a model for finite $n$ as estimated coefficients will not be exactly zero. 

For variable selection, different penalized least squares approaches can be formulated as an optimization problem of the form 
\begin{align}\label{eq:PLS}
\boldsymbol{\hat{\beta}}(\lambda) = \arg \min_{\boldsymbol{\beta}} ||\mathbf{y}-\mathbf{X}\boldsymbol{\beta}||_2^2 + \lambda ||\boldsymbol{\beta}||_q ,
\end{align}
where the penalty term $||\boldsymbol{\beta}||_q:=\left(\sum_{j=1}^{p}|\beta_j|^q \right)^{1/q}$ denotes the $L_q$-norm with special case $||\boldsymbol{\beta}||_0:=\sum_{j=1}^{p}I\left(\beta_j \neq 0 \right)$. The tuning parameter $\lambda \geq 0$ controls the strength of the penalty, where for $q < 2$ a larger $\lambda$ shrinks $\hat{\beta}_j$ more towards 0 and fewer variables are selected. Choosing $\lambda$ can be based on criteria like AIC, BIC, cross-validation or stability procedures \citep{Akaike_1998, Schwarz_1978, Richard_1984, Meinshausen_2010} each with different goals and assumptions \citep{Shao_1997,Yang_2005,Arlot_2010}.
In the following, we will focus on some of the most prominent penalization approaches.

\subsection{Best subset selection}
Using the $L_0$-norm in \eqref{eq:PLS} is known as Best Subset Selection (BSS) and can be formulated as the following discrete optimization problem 
\begin{align}\label{eq:BSS}
\boldsymbol{\hat{\beta}}_{BSS} = \arg \min_{\boldsymbol{\beta}} ||\mathbf{y}-\mathbf{X}\boldsymbol{\beta}||_2^2 \qquad \text{s.t.} \ ||\boldsymbol{\beta}||_0 \leq k
\end{align}
with $k \in \mathbb{N}$ determining the maximal number of non-zeros in $\hat{\boldsymbol{\beta}}$. 
\citet{Zhang_2012} showed that if a uniform signal strength condition for the smallest true predictor in $\boldsymbol{\beta}$ 
holds, BSS can achieve selection consistency. 
\citet{Shen_2013} defined a degree of separation that describes how difficult it is to distinguish the true model from all other models in terms of the projection of $\mathbf{y}$ based on $\boldsymbol{\hat{\beta}}$. As a necessary condition for a $L_0$-norm based penalty approach to be selection consistent, they showed that the degree of separation has to be larger than a threshold that is a function of $p$, $n$ and $\sigma^2$.

The minimization of \eqref{eq:BSS} is known to be NP-hard \citep{Natarajan_1995, Barron_1999} and state-of-the-art algorithms have been capable of solving BSS problems in a feasible amount of time only if $p<50$. Recently \citet{Bertsimas_2016} reformulated \eqref{eq:BSS} as a mixed integer optimization (MIO) problem
\begin{equation}
\begin{aligned}\label{eq:BBS-MIO}
\arg\min_{\boldsymbol{\beta}, \mathbf{z}} &||\mathbf{y}-\mathbf{X}\boldsymbol{\beta}||^2_2 \\
\text{s.t.} \ &(\beta_i,1-z_i) \ : \ \text{SOS-}1, \qquad i=1,\dots,p \\
&z_i \in \{0,1\}, \qquad i=1,\dots,p\\
& \sum_{i=1}^{p}z_i \leq k .
\end{aligned}
\end{equation}
where SOS-1 denotes a Specially Ordered Set of Type 1, i.e., at most one element of $(\beta_i,1-z_i)$ can be non-zero. The authors showed that this reformulation guarantees optimality in the sense of \eqref{eq:BSS}. Due to efficient MIO solvers like {\tt{Gurobi}} \citep{Gurobi}, problem \eqref{eq:BBS-MIO} can be solved in minutes even when $p$ is in the 1000s, $n$ in the 100s and a moderate value $k$ is selected. However, certifying the optimality of the solution can take much more time. For example, the Gurobi solver uses a lower and upper bound criterion to find a solution, where the convergence rate of the lower bound criterion is much faster \citep{Hastie_2020}.

\subsection{Forward step-wise selection}
While BSS is limited through its NP-hard nature, step-wise selection is a popular alternative. It gradually adds (removes) variables to (from) a model based on a model fit criterion. Due to this greedy strategy, these algorithms are computationally less challenging with complexity $\mathcal{O}(p^2)$. However, step-wise selection approaches are known to have numerous drawbacks: they result in unstable final models that are sensitive to small changes in the data \citep{James_1990,Breiman_1996,Whittingham_2006}, and they are only locally optimal, and often miss direct predictors while selecting irrelevant variables \citep{Derksen_1992,Smith_2018}. 
Moreover, inference is problematic as these methods usually do not account for multiple testing issues \citep{Whittingham_2006,Mundry_2009}.
Despite these drawbacks, we will consider forward step-wise selection (FSS) \citep{Efron_2004,Hastie_2009} in our simulation study, because it can be interpreted as greedy heuristic version of BSS \citep{Hastie_2020}. It is defined as an iterative algorithm and starts with an empty active set model $A_0 = \emptyset$ and $\boldsymbol{\hat{\beta}}_{A_0}^{(k)} = \mathbf{0}$. At each step $t=1, \dots,k$, the variable $j_t$ is selected that maximizes
\begin{equation*}
\begin{aligned}
\arg\max_{j_t \notin A_{t-1}} \frac{\mathbf{x}^\top_{j_{t}} (\mathbf{I}-\mathbf{P}_{A_{t-1}})\mathbf{y}}{||(\mathbf{I}-\mathbf{P}_{A_{t-1}}) \mathbf{x}_{j_{t}}||_2}
\end{aligned}
\end{equation*}
where $P_{A_{t-1}}$ denotes the projection of $\mathbf{y}$ onto the column space of $\mathbf{X}_{A_{t-1}}$. Given $j_t$, the active set is updated as $A_t=A_{t-1} \cup \{j_t\}$ and used to estimate 
\begin{equation*}
\begin{aligned}
\boldsymbol{\hat{\beta}}_{A_t}^{(t)} &= \arg \min_{\boldsymbol{\beta}} ||\mathbf{y}-\mathbf{X_{A_t}}\boldsymbol{\beta}||_2^2, \\
\boldsymbol{\hat{\beta}}_{\backslash \{A_t\}}^{(t)} &= 0
\end{aligned}
\end{equation*}
where $\backslash \{A_t\}$ denotes the set of predictors not selected in step $t$.

\subsection{Lasso and Elastic net}
The least absolute shrinkage and selection operator (Lasso) \citep{Tibshirani_1996} uses an $L_1$-norm as penalty term in \eqref{eq:PLS} and shrinks all estimated coefficients towards zero, where some estimates will be exactly zero for a sufficiently large tuning parameter $\lambda$.
	Hence, the Lasso can be used for variable selection as only variables with non-zero coefficients are selected. This property will be used in our simulation study. The total number of zero coefficients is controlled by $\lambda$, where larger values will result in sparser models. The Lasso can be combined with fast algorithms so that problems with $p$ in the $10,000$'s can easily be solved; settings for which BSS with MIO can no longer be applied. However, the Lasso also has several drawbacks. Firstly, it only allows up to $n$ non-zero regression coefficients, which can be a limiting factor if $n \ll p$, \citep{Hastie_2009}. 
Secondly, if irrelevant variables are highly correlated with direct predictors of the outcome $\mathbf{y}$, the Lasso selects almost arbitrarily of those true and false variables and is known not to be consistent, not even for the sign of the coefficient. Furthermore, if there is a high pairwise correlation within a set of variables and they all are true direct predictors for $\mathbf{y}$, the Lasso tends to select only one of these variables \citep{Zou_2005, Xu_2012}. Thus, it cannot guarantee consistent variable selection \citep{Zhao_2006}. To address some of these drawbacks, different modifications of the Lasso have been proposed. They rely on a-priori knowledge about the data generating process or the functional relationship between variables \citep{Tibshirani_2005,Zou_2006,Meinshausen_2007, Friedman_2010,Alaiz_2013,Simon_2013}.

An alternative is the Elastic net (Enet) \citep{Zou_2005}, which can be formulated as a weighted combination of the Lasso and an additional $L_2$-penalization (Ridge) term
\begin{align}\label{eq:enet}
\boldsymbol{\hat{\beta}}_{Enet} = \arg \min_{\boldsymbol{\beta}} ||\mathbf{y}-\mathbf{X}\boldsymbol{\beta}||_2^2 + \alpha \lambda ||\boldsymbol{\beta}||_1 + (1-\alpha) \lambda ||\boldsymbol{\beta}||^2_2.
\end{align}
The second tuning parameter $0 \leq \alpha \leq 1$ controls the weighting between the $L_1$- and $L_2$-penalty.
Here the $L_1$-penalty induces a Lasso-type variable selection, while the $L_2$-penalty helps with highly correlated variables by increasing the diagonal entries of the covariance matrix $\mathbf{X}^\top\mathbf{X}$. The latter guarantees a positive-definite covariance matrix so that it is possible for all $p$ estimated coefficients to be non-zero. More importantly, the $L_2$-penalty can be interpreted as an artificial decorrelation of the variables, making it easier to jointly select highly correlated variables if they are all direct predictors for $\mathbf{y}$ \citep{Zou_2005}.

\section{Simulation study}
\subsection{Simulation design and evaluation for synthetic datasets}
To evaluate the performance of BSS, FSS, Lasso and the Enet for variable selection, we simulated synthetic data from a linear model
(\ref{eq:lin_reg_model})
with $\mathbf{X} \sim \mathcal{N}_p(\mathbf{0}, \boldsymbol{\Sigma})$ and $\mathbf{y} \sim \mathcal{N}_n(\mathbf{X}\boldsymbol{\beta}, \sigma^2\mathbf{I})$. Since real-world applications of variable selection often have a small signal-to-noise ratio $\tau$, we followed \citet{Hastie_2020} and set $\sigma^2=\frac{\boldsymbol{\beta}^\top\boldsymbol{\Sigma}\boldsymbol{\beta}}{\tau}$ with $ 0.05 \leq \tau \leq 6$. 

\begin{figure}[h]
	\centering
	\includegraphics[width=0.97\textwidth]{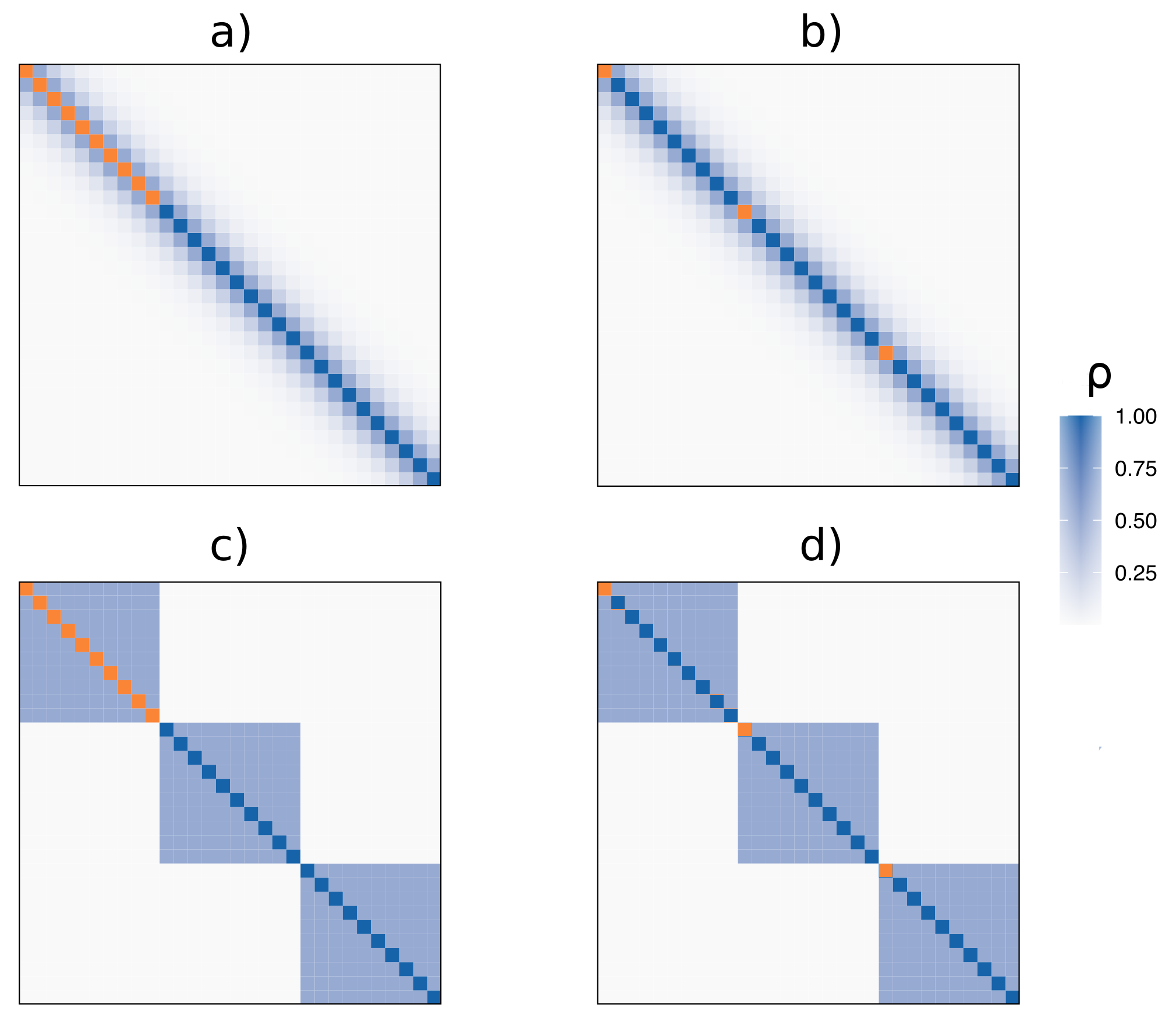}
	\caption{Schematic representation of the different correlation structures and the positioning of the non-zeros coefficients (orange). The Toeplitz structure, with all ten non-zeros in consecutive positioning (a) and equally spaced positioning (b), has a gradually decreasing correlation between all variables. The second row shows a block structure where all coefficients of one block are set to non-zero (c) or only one coefficient of each block is set as non-zero for 10 blocks (d). The correlation between variables within one block is always the same while the correlation between variables of different blocks is set to zero.}
	\label{fig:corr_struc}
\end{figure}

We considered both a low-dimensional ($n=1000$ and $p=100$), high-dimensional ($n=100$ and $p=1000$) and an intermediate ($n=500$ and $n=500$) setting. To assess the effect of correlation between predictors, we used an uncorrelated structure and the standard Toeplitz structure.
The latter was created by setting the pairwise correlation between two variables $x_{\cdot,u}$ and $x_{\cdot,v}$ for $u,v = 1, \dots,p$ as $\rho^{|u-v|}$ with $\rho \in \{0.35, 0.7\}$. Although the Toeplitz structure is a popular correlation structure in simulations, it can be implausible for some applications. For example, in genetic epidemiology, it is often reasonable to assume that genes within a functional group are correlated with each other, but that they are nearly independent of genes from other functional groups. Hence, we simulated data with correlations following a block structure for which variables were grouped into blocks of size $10$. Pairwise correlations within the blocks were set to $\rho \in \{0.35, 0.7\}$; pairwise correlations between variables outside of the blocks were set to $\rho=0$. The number of direct predictors was set to $s=10$ in all settings and their position was either consecutive or equally spaced along the sequence of variables. For the consecutive positioning, we set the first ten coefficients to be non-zeros, while for the equally spaced positioning, we set every tenth coefficient to be non-zero (see Figure \ref{fig:corr_struc}). In all scenarios, a non-zero direct predictor was set to $\beta_j = 1$. Overall, we investigated 270 different scenarios and each one was repeated 100 times.
We used the F1-score 
\begin{align*}
   F_1 = 2 \cdot \frac{P \cdot R}{P + R}
\end{align*}
as main performance measure, where $P=\frac{TP}{TP+FP}$ is the precision and $R=\frac{TP}{TP+FN}$ is the recall, $TP$  the number of true positives, $FP$ the number of false positives, and $FN$ the number of false negatives. Low values reflect that an increase in recall can only be achieved by lowering the precision considerably and vice versa. In this sense, high values reflect a good recall without sacrificing precision. For imbalanced confusion matrices, the F1-score can give misleading results compared to other measures like Matthew's correlation coefficient (MCC)  \citep{Hand_2018,Chicco_2020,Zhu_2020}. However, this is unlikely for high dimensional variable selection problems as the assumed sparsity in these settings implies large numbers of (true) negatives and thus very similar F1-score and MCC values. For the sake of completeness, we still computed the MCC and the F2-score, the latter placing more weight on recall than the F1-score. We found  very similar results; MCC and F2-score results are provide the in the web-app. Further, we investigate the methods' performances for a range of given subset sizes $k$. 

\subsection{Simulation design and evaluation for semi-synthetic datasets}
To assess the performance of the methods in a real-world correlation setting, we generated semi-synthetic data \citep{Buehlmann_2014, Wang2020} using gene expression data from ovarian cancer samples from The Cancer Genome Atlas Program \citep{TCGA2011}. The dataset contains $n=594$ units and $p=22,277$ genes and can be accessed online \citep{Tucker2014,Wang2020}. For the low-dimensional setting, we randomly selected $p=100$ genes and used all $n=594$ observations. For a high-dimensional setting, we randomly selected $p=1000$ genes and $n=100$ observations. The outcome $\mathbf{y}$ was simulated in analogy to the above fully synthetic settings, where we set the signal-to-noise ratio $\tau = 0.42, 1.22, 3.52$ and the coefficient of the 10 direct predictors to $\beta_j = 1$. In determining the direct predictors, we followed the approach of \citet{Buehlmann_2014}, i.e., out of the $p=100$ (or $p= 1000$) variables, we selected the pair with the highest absolute correlation as the first two direct predictors $\mathbf{x}_1$ and $\mathbf{x}_2$. We then chose those eight genes exhibiting the largest correlation with $\mathbf{x}_1$ as further true direct predictors. Each of the semi-synthetic scenarios was repeated 100 times. The empirical mean correlation across pairs of true direct predictors for the low-dimensional setting was $\bar{\rho}\approx0.19$, and $\bar{\rho}\approx 0.37$ for the high-dimensional setting. The methods' performances were again evaluated on the basis of the F1-score. 

\subsection{Selection of tuning parameters} 
All methods under consideration rely either on choosing a tuning parameter $\lambda$ or a set size $k$ a-priori. The Enet further requires two tuning parameters. 
Obviously, the performance of each variable selection method depends on the choice of these tuning parameter(s) or subset size. 
Since we are interested in the best \textit{possible} performance of each method regarding variable selection, we used a grid of tuning parameters $\lambda$ and $\alpha$ for Lasso/Enet, i.e. $\alpha = 0.1, 0.2, \dots, 0.9$ and 1000 values for $\lambda$ were the largest $\lambda$ returns an empty model and the smallest $\lambda$ a full model. For BSS and FSS, we used subset sizes of $k=1,\dots,15$, which means that for $k\geq 10$ both methods had the chance to find the true model. In the final step, we only chose those tuning values/set sizes that gave the highest F1-score (or MCC or F2-score) for that method in the considered setting. We consider this `best possible F1-score' as an indicator of a methods' potential.
In practice, this choice of tuning parameter is not feasible. Different practical approaches for choosing the tuning parameter(s) have been developed, such as the Akaike / Bayesian information criteria  \citep{Akaike_1998, Schwarz_1978}, or (re-)sampling based techniques like cross-validation \citep{Allen_1974, Stone_1974} and stability approaches \citep{Meinshausen_2010, Liu_2010}. However, there is no unique, let alone neutral, standard choice, especially in in high-dimensions and when the aim is to identify the true variables as opposed to mere prediction of the outcome; moreover, re-sampling is computationally intractable with BSS. 
For a more practical and fair assessment, we therefore additionally inspected the performance for each method by choosing the tuning parameters to achieve a {\em given} number of selected variables, which we varied over the range $k=1,2,\dots,15$. This reflects and provides additional insights into the methods' performances when the tuning parameters are set such that a desired number $k$ of variables (subset size) is selected.

Although MIO allows for more efficient BSS, it can potentially still run for hours, making an extensive comparison infeasible without a time limit. Following the suggestion of \citet{Bertsimas_2016} and \citet{Hastie_2020}, we set the time limit for the synthetic data and the semi-synthetic data to 3 minutes and 10 minutes, respectively, for each value of $k$ in each simulation run. These time limits are sufficient for the Gurobi solver to find a solution. However, certifying the solution (proving optimality) can take much longer. As the time limit might disadvantage BSS, we further investigated the number of certified optimal BSS solutions and the impact of varying the time limit on the performance measures in selected settings. In addition, we analyzed the impact of varying the subset sizes $k$ on the BSS performance for the certified solutions.

\section{Results}
Surprisingly, BSS reliably outperformed the other methods only in settings with a high signal-to-noise ratio and when the variables were uncorrelated. 
Even in a \textit{low-dimensional} synthetic data setting, with the number of observations ten times the number of variables, the selection performance of BSS drops dramatically if the true predictors are moderately correlated. In those cases, BSS is even outperformed by the Lasso, which is known to be inconsistent for variable selection when the true predictors are highly correlated. Interestingly, the much simpler FSS achieves similar performance as BSS in almost all settings and, in some, performs even slightly better. The results for the synthetic settings are corroborated by those for the semi-synthetic data settings with the real-world correlation structures. Again, the BSS and FSS performances are similar and, especially in the high-dimensional setting, the performance of BSS does not improve much with a high signal-to-noise ratio. Even in a low-dimensional setting, BSS is clearly inferior to the Enet and the Lasso when the signal-to-noise ratio is low to moderate. The evaluation of the selection performance under different subset sizes supports our approach to use the best possible F1-score for showing the potential of the methods under consideration. In no setting  could we observe an unstable F1-score for an increasing subset size; rather, in correlated settings, BSS and FSS are often incapable of selecting more than a few true direct predictors even for larger subset sizes.

To give some details on the BSS performance, we address a selection of four synthetic data settings. These settings differ with regard to (i) dimensionality, (ii) the position of the non-zero coefficients, (iii) correlation structure, and (iv) correlation strength. We will then report results from the semi-synthetic data settings, and investigate the possible role of the time limit on certification. The remaining results support our main conclusions and are shown as supplementary material or can be seen in the interactive web-app at {\tt{ https://bestsubset.bips.eu}}. The supplementary material and the web-app also show the results for the MCC and the F2-score (both being similar to the results of the F1-score), and the performance with respect to a range of given subset sizes. Note that for the Enet, we only show results with $\alpha \in \{ 0.1, 0.5, 0.9\}$, representing a mostly Ridge-weighted, a balanced and a mostly Lasso-weighted Enet, respectively. All results for the Enet with all nine different $\alpha = 0.1, 0.2, \dots 0.9$ are accessible through our web-app.

\subsection{Variable selection performance}
\begin{figure}[h]
	\centering
	\includegraphics[width=0.9\textwidth]{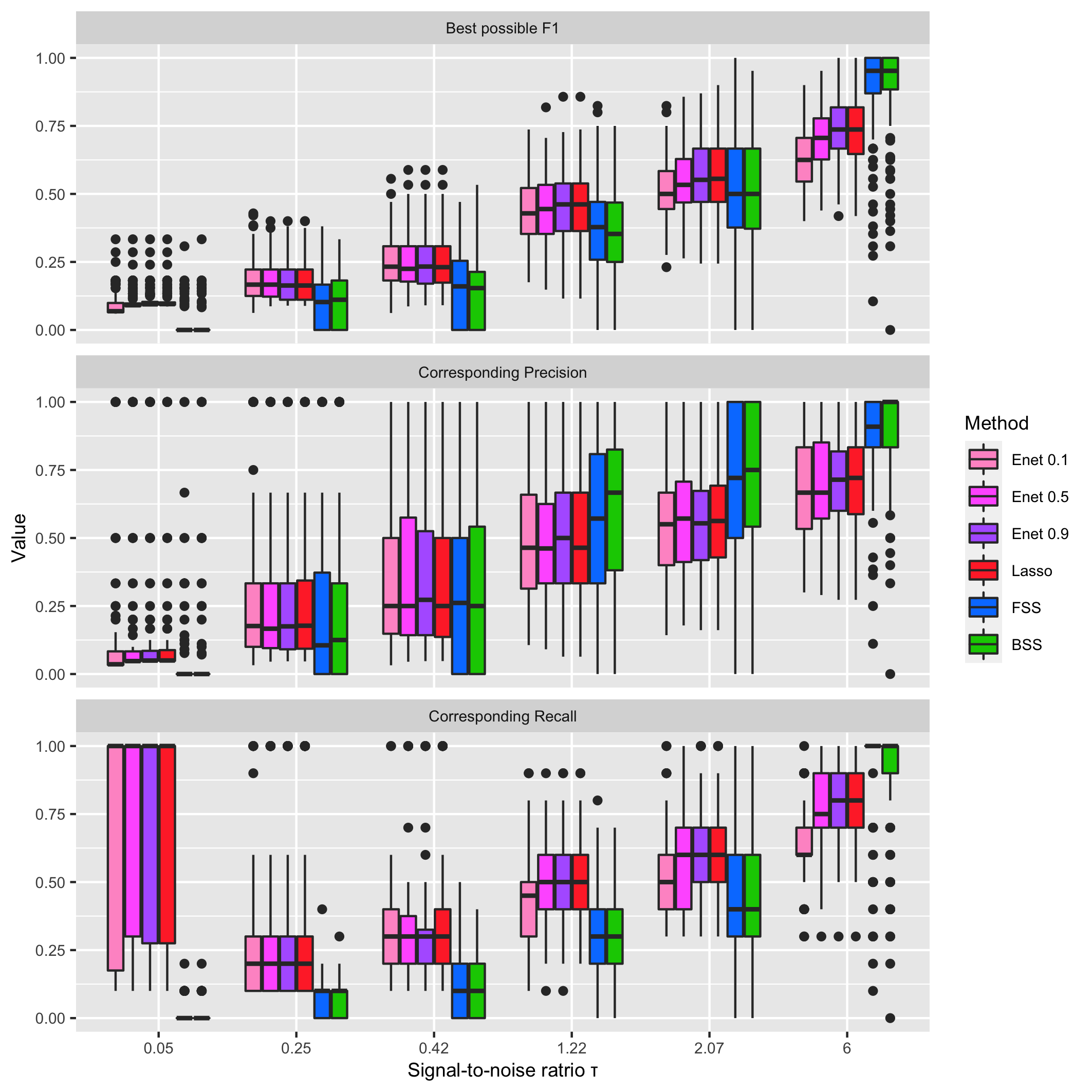}
	\caption{High-dimensional synthetic data setting ($p=1000$, $n=100$, $\rho = 0.35$). Boxplots of best possible F1-scores and corresponding precision and recall values for a block correlation setting with equally spaced non-zero coefficients.}
	\label{fig:blockhighspread35}
\end{figure}

For the synthetic datasets, all methods perform, in general, better with a Toeplitz correlations structure than with a block structure. This seems plausible since the correlations under the Toeplitz structure are weaker than within the blocks. For the high dimensional block setting with equally spaced non-zeros, $\rho = 0.35$ and low $\tau$, all methods have a relatively low best possible F1-score (see Figure \ref{fig:blockhighspread35}). FSS and BSS show nearly identical results and only outperform the Lasso and the Enet when the signal-to-noise ratios are high. 
Lasso and Enet show better corresponding recall values on average except for very high $\tau$ while the variability is high for $\tau < 0.25$. 

\begin{figure}[h]
	\centering
	\includegraphics[width=0.9\textwidth]{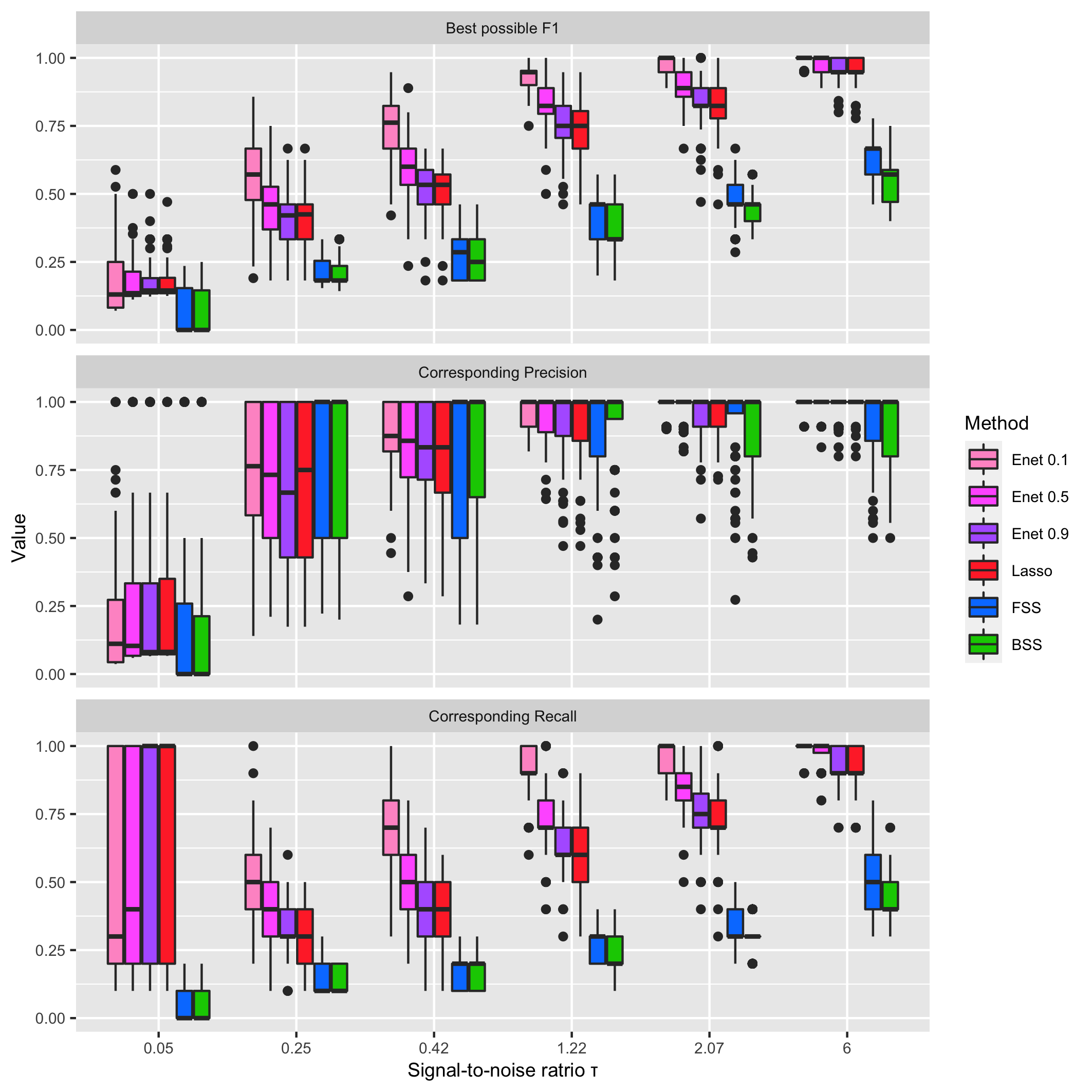}
	\caption{High-dimensional synthetic data setting ($p=1000$, $n=100$, $\rho = 0.7$). Boxplots of best possible F1-scores and corresponding precision and recall values for a Toeplitz correlation setting with consecutive non-zero coefficients.}
	\label{fig:toeplitzhighfirst70}
\end{figure}

Figure \ref{fig:toeplitzhighfirst70} shows the results for a high-dimensional Toeplitz structure setting with consecutive non-zeros. The methods exhibit large differences: for $\tau \geq 2.07$ the Ridge-weighted Enet ($\alpha = 0.1$) performs very well, reaching a high F1-score close to $1$, clearly benefiting from the decorrelation property. In comparison, the other methods cannot cope with highly correlated direct predictors as seen from the low corresponding recall values and low F1-scores of the Lasso, and the weak performances of BSS and FSS. Figure \ref{fig:toeplitzhighfirst70} shows that the recall of BSS and FSS improves slightly with an increase of $\tau$.

\begin{figure}[h]
	\centering
	\includegraphics[width=0.9\textwidth]{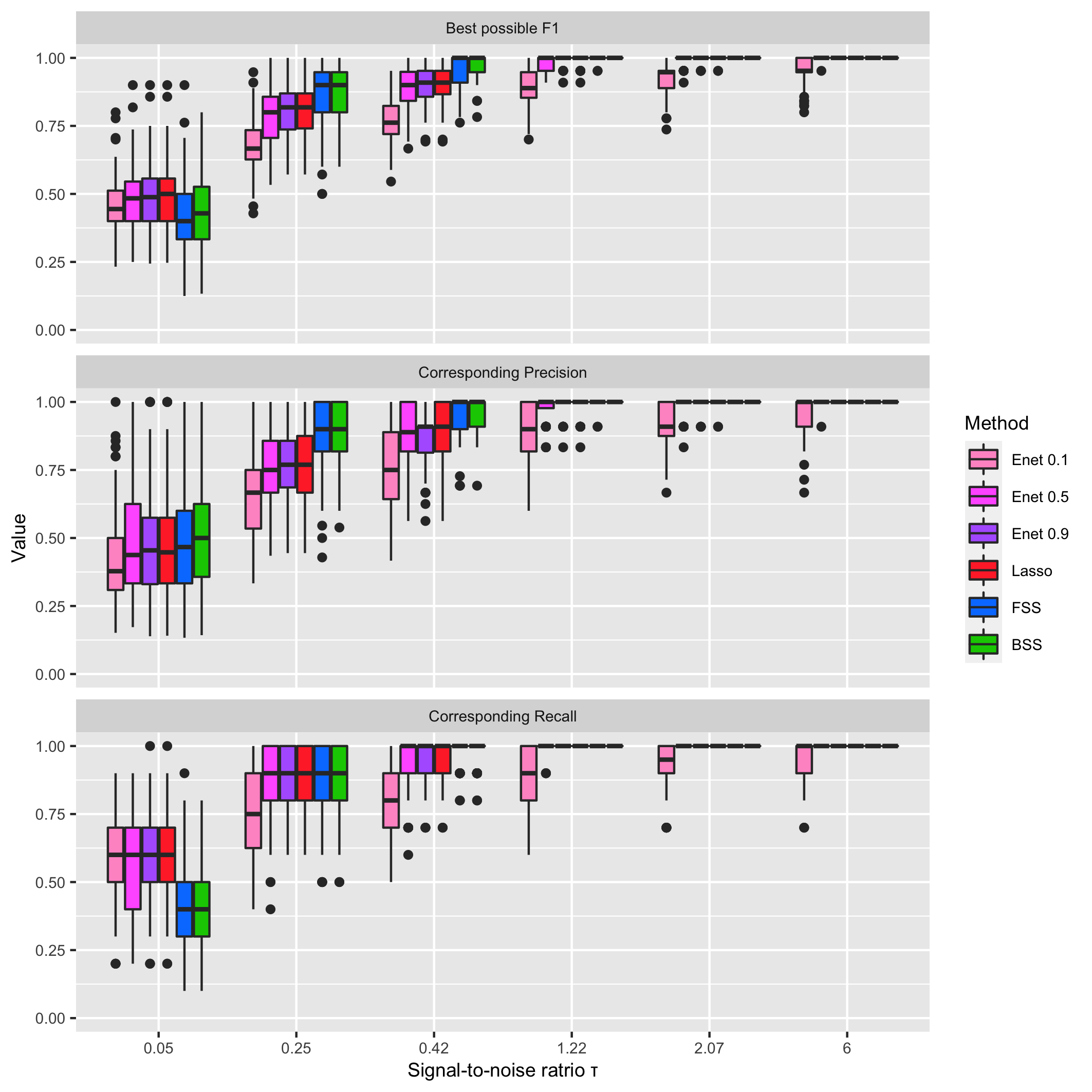}
	\caption{Low-dimensional synthetic data setting ($p=100$, $n=1000$, $\rho = 0.7$). Boxplots of best possible F1-scores and corresponding precision and recall values for a block-structured correlation setting with equally spaced non-zero coefficients.}
	\label{fig:blocklowspread70}
\end{figure}

\begin{figure}[h]
	\centering
	\includegraphics[width=0.9\textwidth]{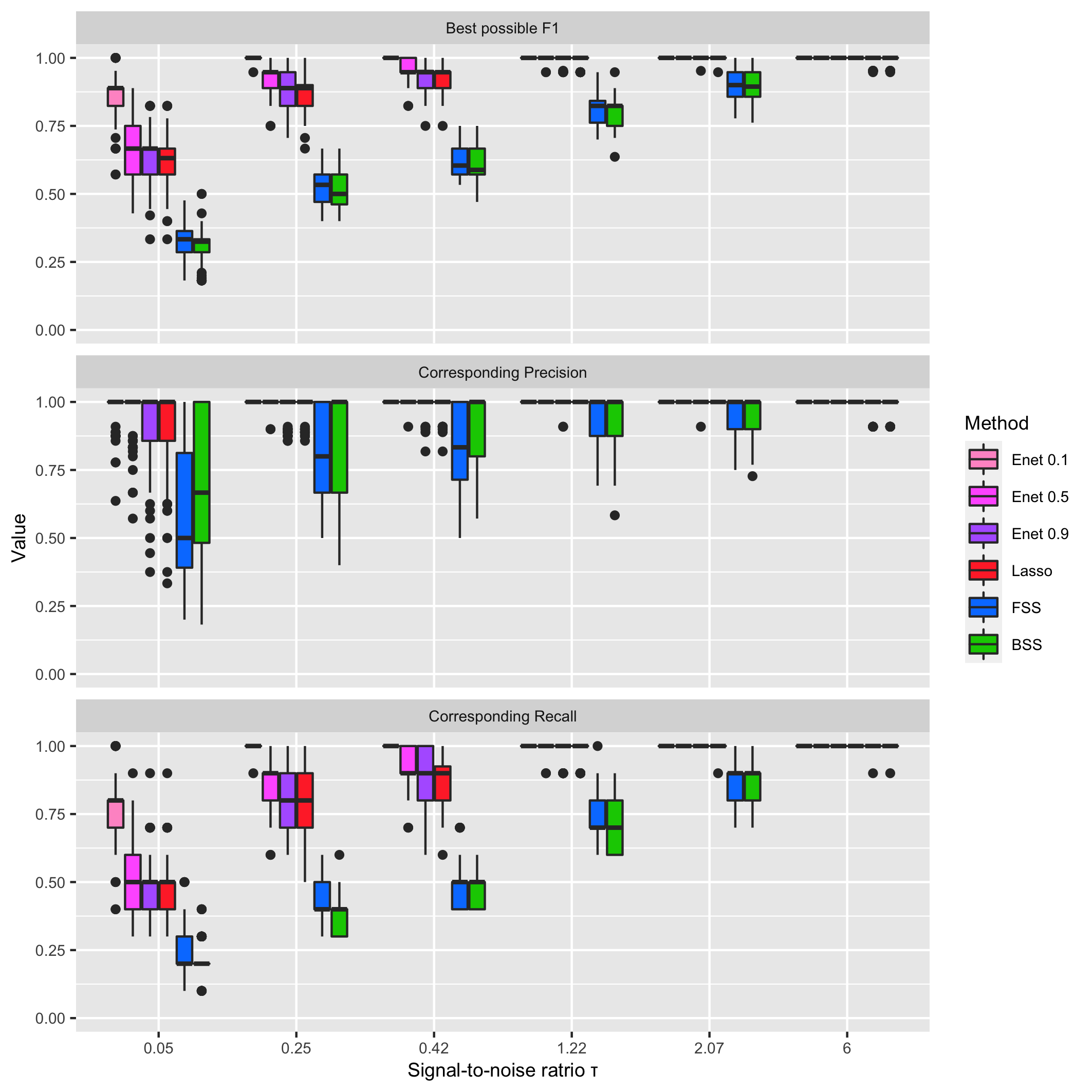}
	\caption{Low-dimensional synthetic data setting ($p=100$, $n=1000$, $\rho = 0.7$). Boxplots of best possible F1-scores and corresponding precision and recall values for a block-structured correlation setting with consecutive non-zero coefficients.}
	\label{fig:blocklowfirst70}
\end{figure}

In the low-dimensional block setting with $\rho = 0.7$ and equally spaced non-zeros, BSS and FSS barely outperform Lasso and Enet for most $\tau$. Figure \ref{fig:blocklowspread70} shows that this is mainly due to the relatively high precision.
However, when the non-zeros are consecutive, the performance of FSS and BSS decreases drastically. Even in the low-dimensional case, the signal-to-noise ratio has to be very large ($\tau = 6$) to achieve comparable results to Lasso and Enet (see Figure \ref{fig:blocklowfirst70}). In these cases, the F1-scores of FSS and BSS are dominated by their poor recall.

\begin{figure}[h]
	\centering
	\includegraphics[width=0.9\textwidth]{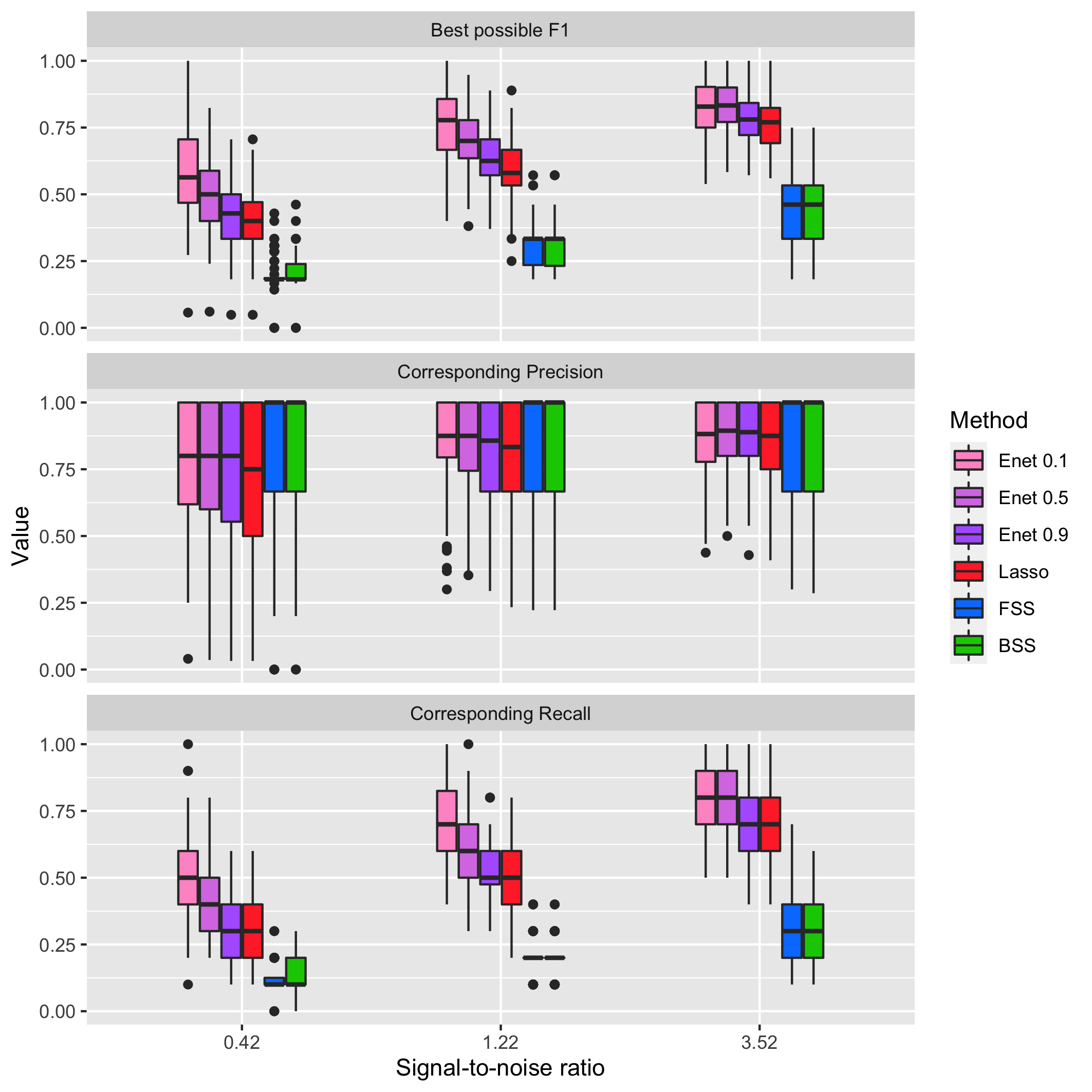}
	\caption{High-dimensional semi-synthetic data setting ($p=1000$, $n=100$) with boxplots of best possible F1-scores and corresponding precision and recall values.}
	\label{fig:semisynhigh}
\end{figure}

\begin{figure}[h]
	\centering
	\includegraphics[width=0.9\textwidth]{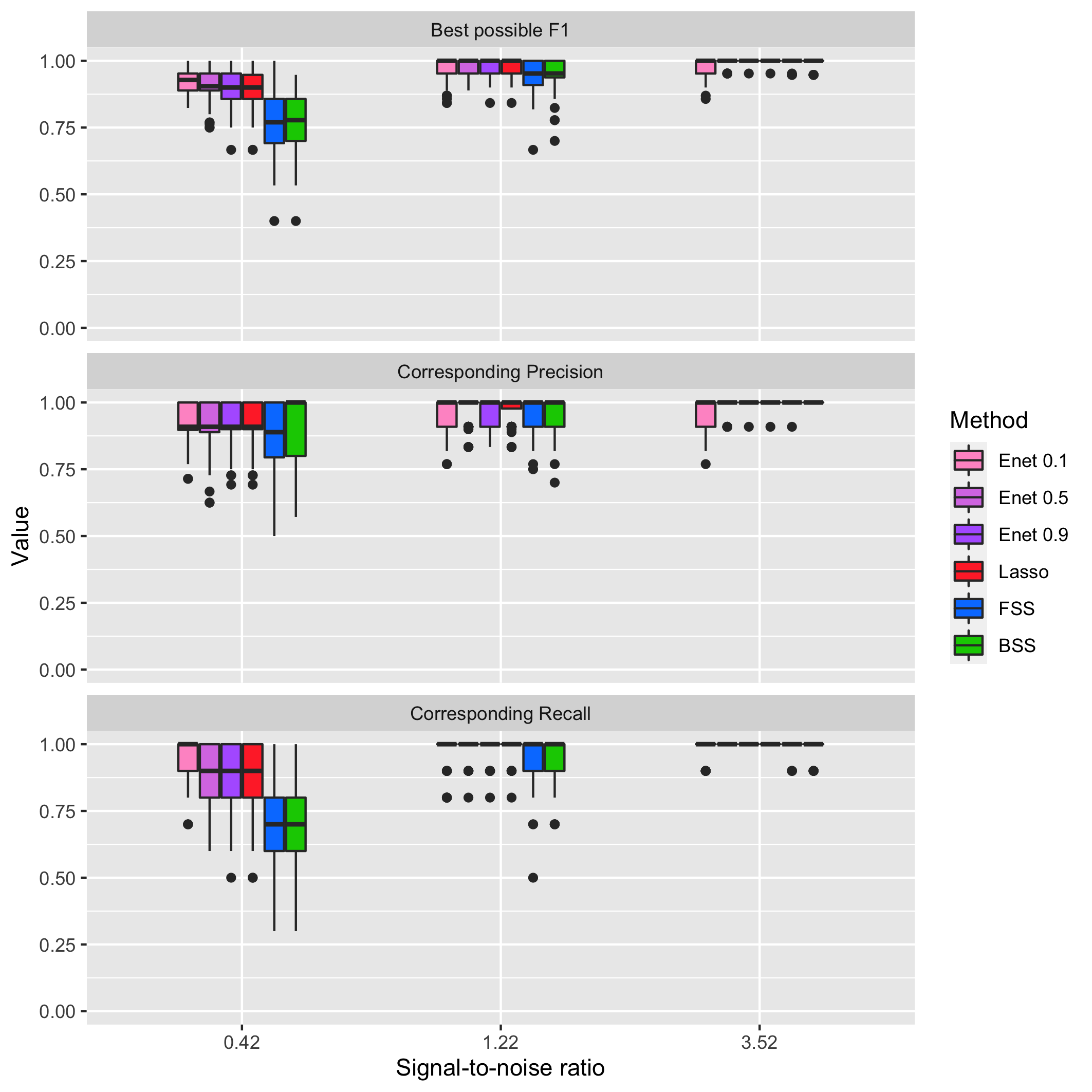}
	\caption{Low-dimensional semi-synthetic data setting ($p=100$, $n=594$) with boxplots of best possible F1-scores and corresponding precision and recall values.}
	\label{fig:semisynlow}
\end{figure}

The results of the semi-synthetic data simulations show a similar tendency. Again, the performances of BSS and FSS are very similar. In the high-dimensional setting (Figure \ref{fig:semisynhigh}), Ridge-weighted Enet versions ($\alpha = 0.1$) achieve an average best possible F1-score $>0.5$ for $\tau=0.42$, while BSS' average best possible F1-score is $< 0.5$
	even for $\tau=3.52$. For the low-dimensional setting (Figure \ref{fig:semisynlow}), BSS is again clearly inferior to Enet and Lasso for $\tau=0.42$ while for $\tau \geq 1.22$ all methods perform very well. The results are similar to the block setting with $\rho=0.35$ for the fully synthetic data simulations (see the supplementary material or the interactive web-app).

\begin{figure}[h]
	\centering
	\includegraphics[width=0.9\textwidth]{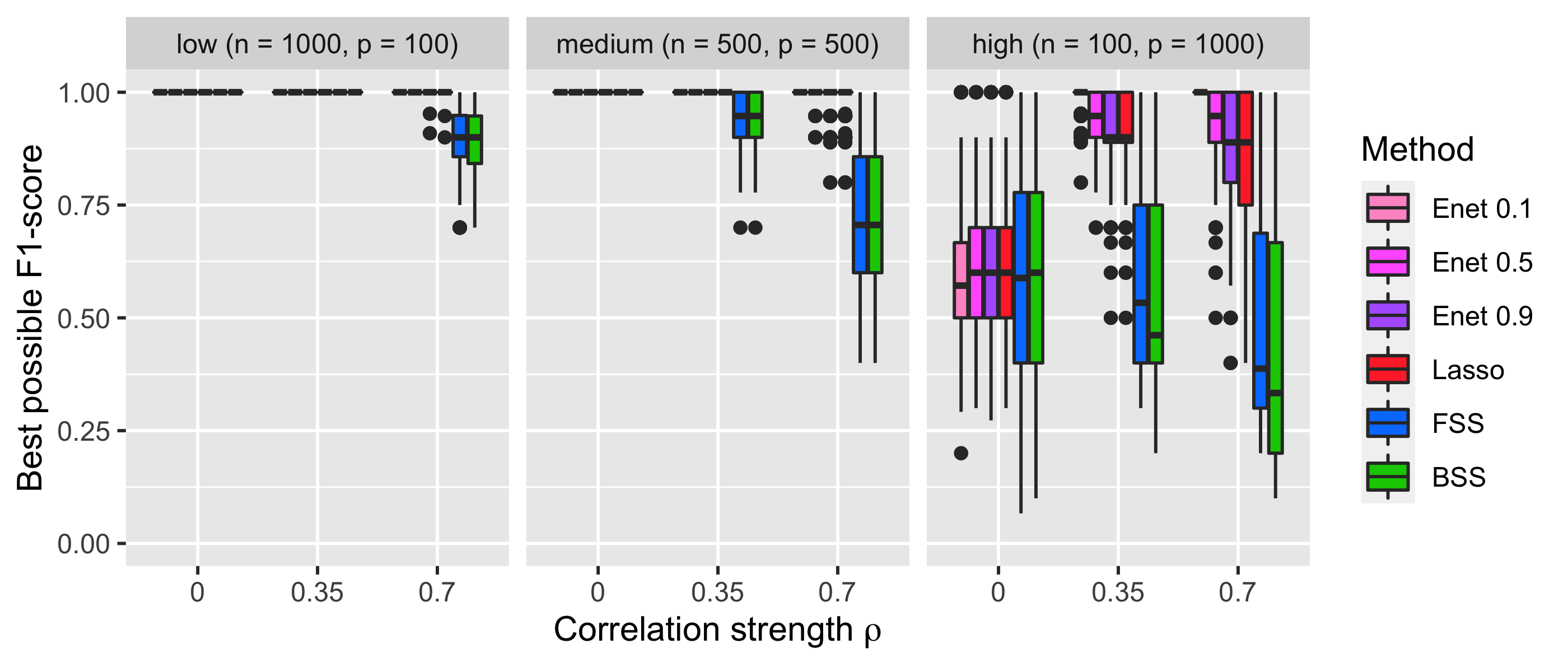}
	\caption{Effect of correlation strength and dimensionality in a block-structured correlation setting with consecutive non-zero coefficients and signal-to-noise ratio $\tau = 2.07$.}
	\label{fig:corr_dim}
\end{figure}

Figure \ref{fig:corr_dim} illustrates the role of the dimensionality and the amount of correlation between true direct predictors on the variable selection performance in a block setting. All methods perform worse in high dimensional settings but they differ strongly in their performance when large correlations are present. While the Enet benefits from the higher correlations due to its grouping effect \citep{Zou_2005,Zhou_2013} Lasso and especially BSS and FSS perform worse. Even in the low-dimensional setting with a relative high signal-to-noise ratio they hardly achieve a perfect F1-score unlike the competitors when there is a large correlation.  

\begin{figure}[h]
	\centering
	\includegraphics[width=0.9\textwidth]{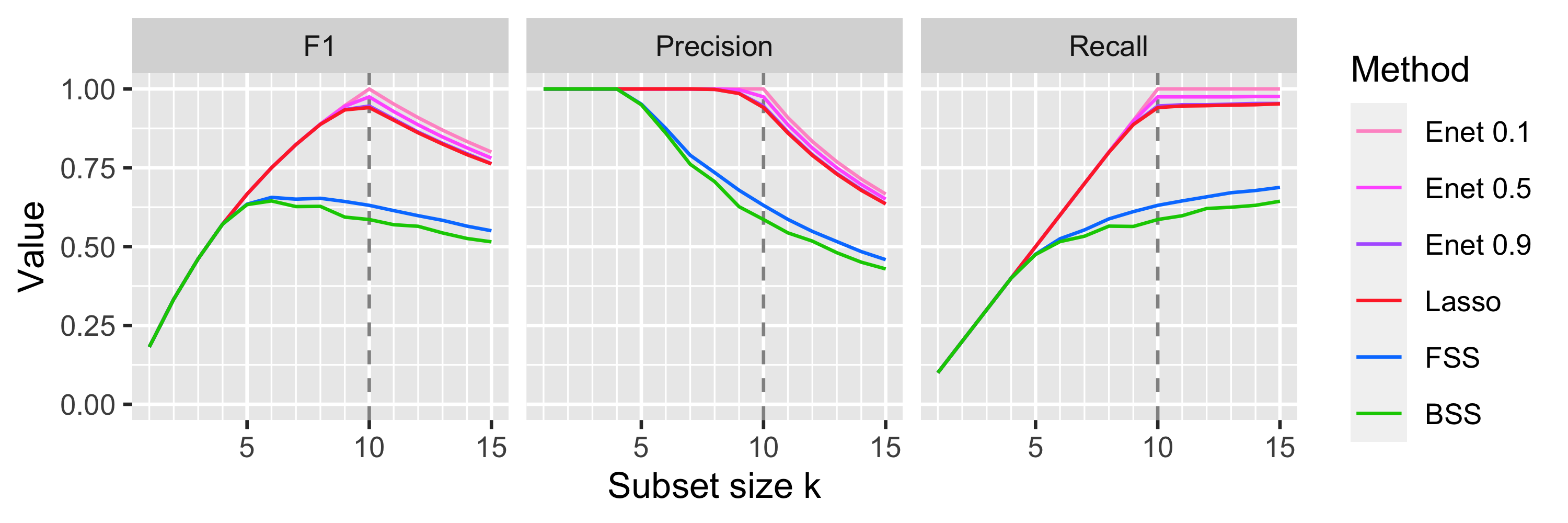}
	\caption{Average variable selection performance based on the subset size $k$ for a low-dimensional block-structured correlation setting ($p=100$, $n=1000$, $\rho = 0.7$, $\tau = 0.71$) with consecutive non-zero coefficients. The dashed vertical line indicated $k=s$, i.e. where the number of selected variables equals the number of true direct predictors.}
	\label{fig:value_k}
\end{figure}

When we choose all tuning parameters to achieve  given subset sizes, we find essentially the same picture as described above regarding the role of correlated variables and signal-to-noise ratios. Moreover, no method exhibited any unstable results (see Figure \ref{fig:value_k} and the web-app). We conclude that slightly sub-optimal tuning parameters or given subset sizes do not cause substantial differences in the performance measures.

\begin{figure}[h!]
	\centering
	\includegraphics[width=0.9\textwidth]{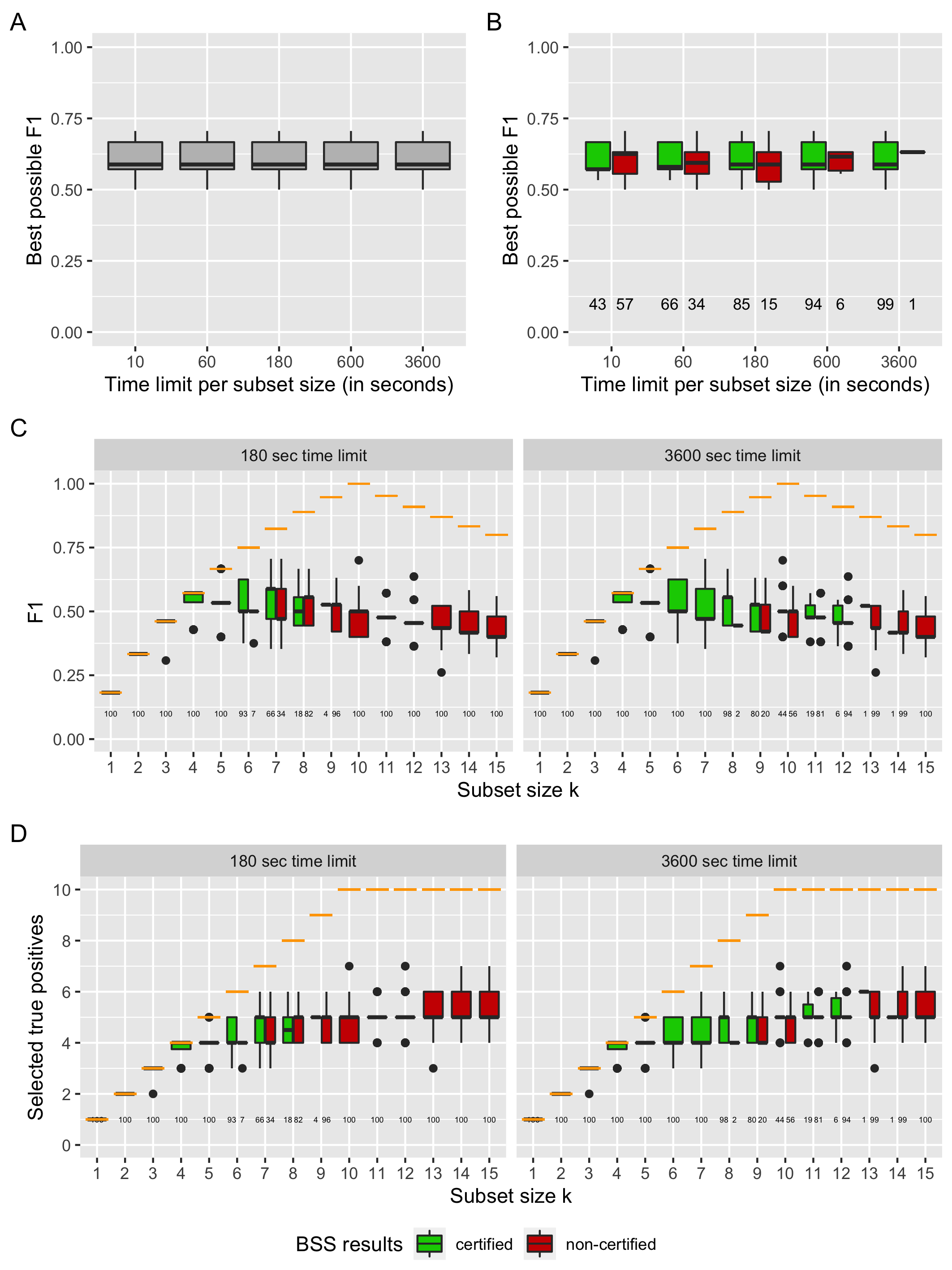}
	\caption{BSS variable selection performance based on different time limits by certification for a block-structured low-dimensional setting with $\rho = 0.7$, consecutive true predictors and $\tau = 0.42$. For all time limits, the same 100 simulated datasets have been used. Panels A and B show the best possible F1-scores across $k$ with respect to the time limit, the latter separately for certified vs. non-certified runs. Panels C and D show the F1-score and the number of selected true direct predictors with respect to the subset size $k$ and the time limit for each subset size. The orange lines in C and D represent the theoretical best value that could be achieved for the given subset size. The numbers underneath the boxplots are the corresponding number of runs for each boxplot.}
	\label{fig:validatedblocklowfirst70}
\end{figure}

\subsection{Time limit, certification of solutions and subset sizes of BSS}
	Although we needed to set a time limit for the Gurobi solver to certify the solutions of each subset size, we conjecture that the poor performance of BSS cannot be explained by non-certified runs or the time limit itself. 
	We investigated this by comparing different time limits, up to one hour, for (the same) $100$ runs in one setting (see Figure \ref{fig:validatedblocklowfirst70}; this is the low-dimensional block setting with consecutive non-zeros, signal-to-noise ratio $\tau=0.42$ and correlation $\rho=0.7$).
	Panels \ref{fig:validatedblocklowfirst70}A and \ref{fig:validatedblocklowfirst70}B suggest that while increasing the time limit leads to a higher proportion of certified runs it does not affect the best possible F1-scores.
However, the F1-score also depends on the specified subset size, with larger subsets taking longer to certify and even one hour possibly being insufficient; 
	too short a time limit might reduce the chances of BSS to identify more direct predictors. 
	Since the number of direct predictors is $s=10$, only a specified subset size of $k=10$ can theoretically result in a perfect F1-score of $1$. Panels \ref{fig:validatedblocklowfirst70}C and \ref{fig:validatedblocklowfirst70}D suggest that neither the time limit nor the certification of the result makes a difference to the performance of BSS. In fact, the BSS methods appear to `hit a ceiling' with selecting (in this particular setting) at most five or six true predictors even when $k\geq 10$ and even when certified. In consequence, the number of false positives necessarily increases to make up the subset size $k$. Note that Enet with $\alpha = 0.1$ achieved for this setting in all runs a perfect F1-score (see Figure \ref{fig:blocklowfirst70}).

We see a similar behaviour of BSS for the low-dimensional semi-synthetic data with $\tau=0.42$ (see Figure 57 supplementary material). As the correlation between the true predictors is weaker ($\bar{\rho} \approx 0.19$) the performance of BSS is generally better. However, regardless of time limit or certification, BSS appears to rarely select all 10 true direct predictors even when $k>10$.

\section{Conclusion and Discussion}
We carried out an extensive simulation study to compare the performance of BSS and its competitors regarding variable selection in a linear regression setting. We investigated a broad range of parameter constellations as well as different and, maybe more realistic correlation structures than previous works have done. Further, we evaluated the methods on semi-synthetic data with actual real-world correlation structures in high- and low-dimensional settings. Our results show that the Enet and the Lasso outperform BSS and FSS in most scenarios. This was unexpected since BSS must have considered the true model as a candidate model in every single run (since $k = 15 > s = 10$).
	Perhaps even more surprisingly, we found that BSS also performs relatively poorly in low-dimensional settings if there is a moderate to high correlation between the true predictors.
BSS outperforms the other methods in the synthetic simulations only under nearly uncorrelated predictors and when the signal-to-noise ratio is very high, which is unrealistic in most practical situations. We also argue that the poor performance of BSS cannot be explained by non-certified runs or time limits.
		Moreover, we detected a weakness of BSS in selecting some but not all
		true direct predictors without adding considerably more false positives; the problem was noticeable in all our simulations settings, not only in the settings investigated for certification, whenever the signal-to-noise ratio was not high or the true direct predictors were correlated (see interactive web-app and supplementary material). This suggests that the objective function (\ref{eq:BSS}) and the discrete nature of the optimisation does not easily allow BSS to distinguish between further true predictors and false ones in such settings. To the best of our knowledge, our work is the first to highlight this shortcoming of BSS. 

In summary, the $L_0$-norm penalization seems only appropriate in situations where a high signal-to-noise ratio and (nearly) uncorrelated true predictors are plausible. Alternatively, based on our empirical results, using a Ridge-weighted Enet seems a good choice for settings with correlated predictors. 

Our complete set of results can be accessed via a web-app at {\tt{ https://bestsubset.bips.eu}} and can be found in the supplementary material. We would like to reiterate that our simulation study was designed to evaluate the variable selection methods, and not to assess the criteria for selecting the tuning parameters or subset size. In addition, we focused on scenarios with non-zero predictors of the same positive size. It is reasonable to assume that different sizes and signs will alter the performances. However, there is no reason to assume that this  would specifically improve the performance of BSS to the other methods.

In future research, it might be promising to combine BSS or FSS with approaches like Lasso or Enet so as to preselect or decorrelate covariates. Hence, further insights on the role of the correlation structure for variable selection performance of BSS are desirable. This may not only resolve the computational complexity of BSS but also its poor performance with correlated predictors.

\subsubsection*{Acknowledgements}
We acknowledge financial support by the Deutsche Forschungsgemeinschaft (DFG)  through project FO 1045/2-1.

\subsubsection*{Conflict of Interest}
\noindent {The authors have declared no conflict of interest.}

\subsubsection*{Data Availability}
\noindent {All (semi)-synthetic data of this simulation study can be generated by the R-code under 
\begin{itemize}
    \item {\tt{https://github.com/bips-hb/bsscomparison}},
    \item {\tt{https://github.com/bips-hb/simsham}} and
    \item {\tt{https://github.com/bips-hb/semisynthetic\_data\_simulation}}.
\end{itemize}
The TCGA data for generating the semi-synthetic data simulation is available from the authors website 
\begin{center}
    {\tt{https://bioinformatics.mdanderson.org/Supplements/ResidualDisease/}}
\end{center}  
All results of the simulation can be accessed under {\tt{https://bestsubset.bips.eu}}. \\
All raw results of the medium- and high-dimensional synthetic settings can be downloaded from 
\begin{center}
{\tt{https://www.bips-institut.de/fileadmin/downloads/BestSubsetResults.zip}} 
\end{center}
Raw data of all the semi-synthetic and low-dimensional synthetic settings are stored in in the repository {\tt{https://github.com/bips-hb/bsscomparison}}.

\bibliography{Choosing_the_best_subset} 

\end{document}